\documentclass[onecolumn,11pt]{article}
\usepackage[top=1in, bottom=1in, left=1in, right=1in]{geometry}
\usepackage{amsmath,amsfonts,amscd,amssymb}
\usepackage{graphicx}
\usepackage{epstopdf}
\usepackage{overpic}
\usepackage{cancel}
\usepackage{rotating}
\usepackage{url}
\usepackage{caption}
\usepackage{color}
\usepackage{rotating}
\usepackage{multirow}
\usepackage{wrapfig}
\usepackage{mathtools}
\mathtoolsset{showonlyrefs}
\usepackage{physics}
\usepackage{subeqnarray}
\usepackage{setspace}
\usepackage{subcaption}
\usepackage{palatino} 
\usepackage[numbers,sort&compress]{natbib}
\usepackage[bottom,flushmargin,hang,multiple]{footmisc}
\usepackage{lipsum}
\newcommand\blfootnote[1]{%
  \begingroup
  \renewcommand\thefootnote{}\footnote{#1}%
  \addtocounter{footnote}{-1}%
  \endgroup
}

\definecolor{header1}{cmyk}{0,0,0,1}

\graphicspath{{figures/}}

\newcommand{\R}{\mathbb{R}}
\newcommand{\x}{\mathbf{x}}

\newcommand*{\vertbar}{\rule[-1ex]{0.5pt}{2.5ex}}

\title{\vspace{-.55in}{\fontsize{16}{16}\selectfont \textbf{Physics-informed machine learning\\ for sensor fault detection with flight test data}}\vspace{-.15in}}

\author{\normalsize{Brian M. de Silva$^{1*}$, Jared Callaham$^2$, Jonathan Jonker$^3$, Nicholas Goebel$^4$, Jennifer Klemisch$^4$,}\\
\normalsize{Darren McDonald$^4$, Nathan Hicks$^4$, J. Nathan Kutz$^1$, Steven L. Brunton$^2$, Aleksandr Y. Aravkin$^1$}\\
\footnotesize{$^1$ Department of Applied Mathematics, University of Washington, Seattle, WA 98195, United States}\\
\footnotesize{$^2$ Department of Mechanical Engineering, University of Washington, Seattle, WA 98195, United States}\\
\footnotesize{$^3$ Department of Mathematics, University of Washington, Seattle, WA 98195, United States}\\
\footnotesize{$^4$ The Boeing Company, Seattle, WA 98008, United States\vspace{-.1in}}
}
\date{}
\begin{document}
\maketitle

\blfootnote{$^*$ Corresponding author (bdesilva@uw.edu).}

\vspace{-.2in}
\begin{abstract}
  We develop data-driven algorithms to \textit{fully automate} sensor fault detection in systems governed by underlying physics.
  The proposed machine learning method uses a time series of \textit{typical} behavior to approximate the evolution of measurements of interest by a linear time-invariant system.  
  Given additional data from related sensors, a Kalman observer is used to maintain a separate real-time estimate of the measurement of interest.
  Sustained deviation between the measurements and the estimate is used to detect anomalous behavior.
  A decision tree, informed by integrating other sensor measurement values, is used to determine the amount of deviation required to identify a sensor fault.
  We validate the method by applying it to three test systems exhibiting various types of sensor faults: commercial flight test data, an unsteady aerodynamics model with dynamic stall, and a model for longitudinal flight dynamics forced by atmospheric turbulence.
  In the latter two cases we test fault detection for several prototypical failure modes.
  The combination of a learned dynamical model with the automated decision tree accurately detects sensor faults in each case.

\vspace{0.15in}
\end{abstract}

\section{Introduction}
  Sensor fault detection is an important problem in many fields of engineering, where monitoring and state estimation are required for a system's successful operation.
  Although catastrophic failures may be obvious, more insidious fault modes,  such as decalibration, slow drift, and low-frequency oscillations, are more difficult to detect. 
  Thus, there is a reliance in practice on engineering expertise and heuristics for anomaly identification.
  In the context of aircraft dynamics, redundant measurements are typically taken for critical quantities; a degree of robustness to faults is therefore achieved via a voting or weighted averaging sensor fusion scheme~\cite{allerton2005review}.
  This is not always the case in flight test scenarios, where such large quantities of data are collected that it is impractical to either automatically detect faults via redundancy or to manually monitor them.
  Fast, automatic anomaly detection can lead to reduced flight test time, resulting in significant cost savings for aircraft programs. 
  Moreover, in routine operations, current designs require redundant sensing and flight controllers that are robust to faulty measurements.  
  Robust and guaranteed fault detection may lead to designs with improved performance and reduced environmental impact~\cite{Goupil2011addsafe}.

  Automated model-based fault detection has been well-studied for linear systems~\cite{Mehra1971, Willsky1976, Venkatasubramanian2003}, but the general problem remains open for nonlinear dynamics~\cite{Surana2018jns}.
  When a nonlinear model is available, one approach to detect anomalous sensor behavior is to 
  estimate the state with an extended Kalman filter~\cite{ljung1979asymptotic,welch1995introduction}, and to then detect consistent discrepancies between the model and estimates by comparing 
  observed deviations to an expected range of process noise~\cite{Li1991, Walker1994, delGobbo2001}.
  This approach has been applied to flight dynamics models for identification~\cite{Hajiyev2010} and automatic isolation~\cite{vanEykeren2014} of anomalous sensors.

  If a physics-based model is unavailable, data-driven methods offer an attractive alternative to detect faults, for example by training neural networks~\cite{Napolitano1993, Raza1994, Napolitano1996}.
  Although in recent years neural networks have had impressive successes in fields such as image and speech recognition~\cite{bengio2017deep}, their behavior tends to be unreliable outside of training conditions: the precise regime that is arguably most important for investigating anomalous behavior.
  Indeed, deep learning models are well known to consistently fail in generalization and extrapolation tasks since they are interpolatory in nature~\cite{mallat2016understanding}.
  An alternative is data-driven system identification~\cite{ERA:1985, Juang1991, Schmidt2009science, schmid_dynamic_2010, Brunton2013jfm,Tu2014jcd,Hemati2016aiaa, brunton_discovering_2016,Brunton2019book,Brunton2020arfm}.
  These methods can identify a dynamical model that is suitable for filter-based state estimation and fault detection~\cite{Surana2016, Surana2018jns}.

  In this paper we build on previous work in data-driven model identification and anomaly detection, in particular~\cite{vanEykeren2014} and~\cite{Surana2018jns}, by combining optimal estimation and system identification with modern machine learning methods.
  We verify that the proposed approach is applicable to both strongly nonlinear dynamics and correlated, non-Gaussian process noise, and demonstrate fault detection on data from flight tests.
  A core component of the algorithm is a simple Kalman observer which is used to predict future sensor values based on current measurements.
  When the difference between this prediction and the observed values passes a threshold over a period of time, a sensor fault is flagged. 
  The Kalman filter requires a model of the underlying dynamics from which to estimate future states.
  This model is learned from data via the {\em dynamic mode decomposition} (DMD)~\cite{schmid_dynamic_2010,kutz_dynamic_2016} with time-delays~\cite{brunton_chaos_2017}.
  A decision tree~\cite{safavian1991survey} is then used to determine the right threshold for the gap between prediction and observation.
  The model is fully automatic; one need only specify a set of labeled training data and it will learn both a model for the dynamics present in the data and a set of rules for detecting sensor failures.

  The key advantages of this approach are that it is fully data-driven, so that  a model is not required for the dynamics underlying the system being monitored, only measurements.   It is also automated and can readily support a large number of measurements/features.
  Its primary disadvantage is that it is a supervised method, meaning that one must supply labeled data to train the model.
  The techniques from which the model is built are all fairly general, granting it a large amount of flexibility while simultaneously restricting its accuracy for some specific applications.
  In domains in which underlying physical dynamics are well-understood, it may be advantageous to use a more specialized method.

  The paper proceeds as follows. In Section \ref{sec:background} we give an overview of the mathematical background underlying our approach before describing the proposed method itself in Section \ref{sec:proposed-method}. Section \ref{sec:examples} discusses three example flight applications: one real-world flight test dataset and two simulated examples. We conclude with Section \ref{sec:conclusion} which provides some final thoughts.

\section{Background}
\label{sec:background}

  In this section we provide a mathematical foundation for the proposed method.
  Sections \ref{sec:kalman-filter} and \ref{sec:kalman-detection} describe Kalman filters and their application to anomaly detection.
  A method for constructing a linear time invariant (LTI) model for use in a Kalman filter is discussed in Sections \ref{sec:dmd} and \ref{sec:dmdc}.
  Section \ref{sec:dmd-hankel} details a method of enriching the LTI model by introducing time-delays.
  Finally, Section \ref{sec:decision-tree} gives a brief discussion of decision trees, the last component of our method.

  \subsection{System identification and Kalman filtering}
  \label{sec:kalman-filter}

    Since their development in the 1960's~\cite{Kalman1960jfe}, various forms of Kalman filters have proven useful in fields ranging from robotics to weather prediction.  The filter described in this section is a simple form of this powerful tool, but is nonetheless effective in many test problems.  The method is essentially a simplification of those proposed in~\cite{delGobbo2001, Hajiyev2010, vanEykeren2014, Surana2018jns}, and references therein. It is thus potentially extensible to more complex detection and estimation problems, including those with underlying physical systems exhibiting strongly nonlinear dynamics.

    Although application of Kalman filters to fault detection has been proposed since the 1970's, until recently an existing model was necessary for the method.  Considering the scale of sensing in flight test applications, developing independent predictive models for the various sensors could be prohibitive.  However, a recent development suggested in~\cite{Surana2018jns} was to identify a linear time-invariant (LTI) model that estimates the relationship between measurements using the DMD algorithm.  DMD is a powerful method originally developed in the fluids community to study spatio-temporal coherence in high-dimensional numerical and experimental fluid flow data~\cite{schmid_dynamic_2010, Rowley2009jfm, Tu2014jcd}.  It has since found applications ranging from neuroscience to epidemiology~\cite{kutz_dynamic_2016}.  The method is designed to efficiently extract dominant patterns from very large data sets and automatically uses correlations in the data for improved robustness.

    A simple linear model identified with this algorithm may not be accurate enough to account for the complex interactions in the aircraft system.  However, although Kalman filters are often used in full-state estimation, if the goal is not estimation but fault detection, the dynamic model does not need to be particularly accurate in order to capture anomalous behavior, as our results demonstrate.  The first step of the procedure is therefore to identify a linear predictive model from a time series of {\em typical} measurements.  Online, this DMD model is used to maintain a Kalman filtered estimate of the measurement of interest.  The variance between the estimate and the actual measurement is monitored, and persistent deviations signal anomalous behavior.

  \subsection{Anomaly detection with Kalman filters}
  \label{sec:kalman-detection}

\begin{figure}
    \centering
    \includegraphics[width=\textwidth]{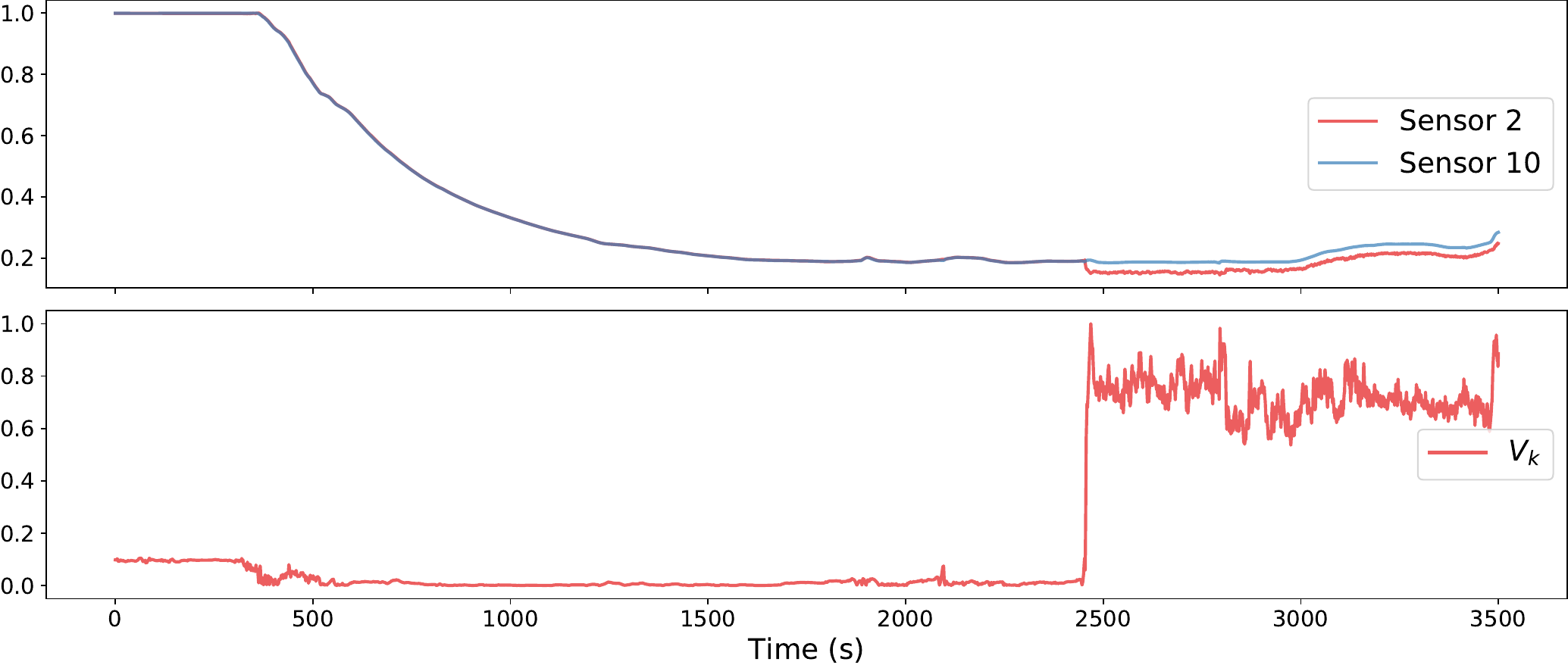}
    \caption{Anomaly detection for the flight test dataset of Section \ref{sec:real-world-dataset} with the Kalman filter. Top: measurements from two redundant sensors. Just before 2500 seconds, sensor 2 breaks and begins giving erratic readings. Bottom: a moving average, $V_k$, of the covariance term (see \ref{eq:avg-innovation-cov}). Note that $V_k$ remains negligible until the sensor failure event leads to persistent anomalous measurements relative to the Kalman filter.}
    \label{fig:example-flight-covariance}
    \end{figure}

    The following gives a simplified description of what will be used in our approach outlined in Section \ref{sec:proposed-method}.   More sophisticated methods can be found in~\cite{delGobbo2001, Hajiyev2010, vanEykeren2014, Surana2018jns}.   Let the vector of measurements we wish to monitor at discrete time step $ k $ be denoted by $ \mathbf{x}_k \in \R^n.$  If we have access to a set of related (but not necessarily redundant) measurements, denote these by $ \mathbf{y}_k \in \R^p.  $  For the sensor fault detection examples, $ x, $ a scalar, is the sensor measurement, and $ \mathbf{y} $ is a vector of other relevant measurements (e.g. readings from other sensors).  Assume we have an LTI model (identified either by dynamic mode decomposition or some other procedure) that predicts the next measurement $ \mathbf{x}_{k+1}, $ given current information $ \mathbf{x}_k $ and $ \mathbf{y}_k $:
    \begin{equation}
      \mathbf{x}_{k+1} = \mathbf{A}\mathbf{x}_k + \mathbf{B}\mathbf{y}_k.
    \end{equation}
    In other words, we treat the related measurements as exogenous inputs in the model.  There does not need to be a direct causal relationship in the sense that actuation is usually taken in control theory; these measurements should just help to predict the next measurement of interest.  Again, this model may be fairly inaccurate.  The use of exogenous inputs is designed to stabilize the model and help to detect drift, high frequency noise, etc.  This model may alternately be viewed as a linear regression predicting the next measurement.   The simplest Kalman filter is a separate LTI ``observer'' system that maintains an estimate $ \mathbf{\hat{x}} $ of the measurement of interest:
    \begin{equation}\label{eq:kalman-filter}
      \mathbf{\hat{x}}_{k+1} = \mathbf{A}\mathbf{\hat{x}}_k + \mathbf{B}\mathbf{y}_k + \mathbf{K}(\mathbf{x}_k - \mathbf{\hat{x}}_k).
    \end{equation}
    The new term in this equation is the innovation $ (\mathbf{x} - \mathbf{\hat{x}}) $ which acts as feedback to stabilize the estimate.  In general the Kalman gain $ \mathbf{K} $ is a matrix, which can be chosen for optimal convergence of the estimate to the true state (by solving a Ricatti equation), given knowledge of the sensor and process noise covariances.  In the example of Fig.~\ref{fig:example-flight-covariance} (and for the numerical examples of Section \ref{sec:examples}), $K$ is a scalar affecting the time sensitivity of the estimated state. We found that the method is relatively insensitive to the choice of Kalman gain, and values from $0.1$ to $0.001$ were tested with similar performance.

    As suggested in~\cite{Mehra1971, Hajiyev2010}, a moving average of the innovation covariance can be used to identify anomalous behavior:
    \begin{equation}\label{eq:avg-innovation-cov}
      V_k = \frac{1}{N}\sum_{i=k-N}^{k} (\mathbf{x}_i - \mathbf{\hat{x}}_i)(\mathbf{x}_i - \mathbf{\hat{x}}_i)^\top.
    \end{equation}
    Intuitively, this term will remain large when a measurement persistently yields anomalous behavior, provided the LTI model captures the important correlations between sensors.
    Even if the model is not particularly accurate, exogenous inputs can still produce an estimate that reflects anomalous behavior.  It is worth emphasizing, however, that without significant additional validation there is no reason to believe that the Kalman filtered estimate is accurate.

    For example, Fig. \ref{fig:example-flight-covariance} demonstrates the Kalman filter-based detection of a sensor failure in the real world data set (see Section \ref{sec:real-world-dataset}).
    An anomaly can be flagged when the innovation covariance exceeds some threshold, which may itself be selected in an automated fashion.
    Although the filtered estimate is often inconsistent with both the faulty sensor and a redundant, working sensor, before and after the faulty sensor fails, the innovation covariance only grows significantly after the failure.
    In other words, the filtered estimate does not need to be accurate in order to be an accurate predictor of sensor faults.

  \subsection{Identifying a linear time-invariant model}
  \label{sec:dmd}

    The method described above requires a reasonably accurate model of the dynamics of the measurement of interest, possibly including the relationship between this measurement and the output of related sensors.  In the simplest case (presented here) this could be a linear time-invariant (LTI) system, although application to a nonlinear model is possible using an Extended Kalman Filter (EKF)~\cite{delGobbo2001, Hajiyev2010, vanEykeren2014} or optimization-based nonlinear Kalman smoothing approaches~\cite{aravkin2012robust,aravkin2014optimization}. 

    Although obtaining a model in general can be a labor-intensive and problem-specific task, recent developments in system identification have enabled a range of straightforward, efficient model estimation tools.  The method presented here uses one such system identification algorithm:  DMD.  DMD was originally developed in the fluid dynamics community as a method to extract coherent spatio-temporal structures from complex, high-dimensional data~\cite{schmid_dynamic_2010,Rowley2009jfm, Tu2014jcd}.  As such, it is designed to take advantage of correlations in the data and reduce the underlying dimensionality of the model.  Although there have been many theoretical and numerical refinements of DMD proposed (see e.g.~\cite{Jovanovic2014, Proctor2016siam,williams_datadriven_2015, williams_kernel-based_2015, brunton_chaos_2017,nathan2018applied,askham_variable_2018}), we present a simple formulation of the method in the paragraphs that follow.

    Suppose we have a series of measurements $ \left\{ \mathbf{x}_1, \mathbf{x}_2, \dots, \mathbf{x}_{m+1}  \right\} $ and we assume that these are related by approximately linear dynamics, i.e.
    \begin{equation}
      \mathbf{x}_{k+1} \approx \mathbf{A} \mathbf{x}_{k}.
    \end{equation} 
    If we arrange the measurements in a time-shifted pair of matrices $ \mathbf{X} $ and $ \mathbf{X'} $ so that
    \begin{equation}
      \mathbf{X} = \begin{bmatrix}
        \vertbar & \vertbar & & \vertbar \\
        \mathbf{x}_1 & \mathbf{x}_2 & \cdots & \mathbf{x}_m\\
        \vertbar & \vertbar & & \vertbar 
      \end{bmatrix}, \hspace{2cm}
      \mathbf{X}' = \begin{bmatrix}
        \vertbar & \vertbar & & \vertbar \\
        \mathbf{x}_2 & \mathbf{x}_3 & \cdots & \mathbf{x}_{m+1}\\
        \vertbar & \vertbar & & \vertbar 
      \end{bmatrix},
    \end{equation}
    these matrices are related by
    \begin{equation}
      \mathbf{X}' \approx \mathbf{A} \mathbf{X}.
    \end{equation}

    Denoting the pseudoinverse of $ \mathbf{X}$ by $ \mathbf{X}^\dagger$,  the least-squares estimate of $ \mathbf{A} $ is given by
    \begin{equation} \label{eq:dmd-pinv}
      \mathbf{A} \approx \mathbf{X}' \mathbf{X}^\dagger.
    \end{equation}
    The spectral properties of the system are then estimated as usual by the eigendecomposition of $ \mathbf{A}. $

    For the case of high-dimensional systems, significant computational gains can be realized by reducing the dimensionality of the problem.  The rank of $ \mathbf{A} $ is limited by the minimum dimension of $ \mathbf{X}$ and  $\mathbf{X}'$.  Instead of studying the spectral properties of the full-state system, we can project the high-dimensional state onto the leading principal components of $ \mathbf{X} $ and approximate the spectrum of $ \mathbf{A} $ by the spectrum of the matrix that steps this low-dimensional approximation forward in time.  That is, if the singular value decomposition of $ \mathbf{X} $ is given by
    \begin{equation}
      \mathbf{X} = \mathbf{\Psi} \boldsymbol{\Sigma} \mathbf{V}^*,
    \end{equation} then the projection of an arbitrary snapshot $ \mathbf{x}_k $ onto the leading $ r $ principal components is
    \begin{equation}
      \alpha_k =\mathbf{\Psi}_r^* \mathbf{x}_k,
    \end{equation} where $ \mathbf{\Psi}_r $ consists of the first $ r $ columns of $ \mathbf{\Psi}. $
    The spectrum of $ \mathbf{A} $ can be approximated by the spectrum of $ \mathbf{\tilde{A}}, $ where
    \begin{equation}
      \alpha_{k+1} \approx \tilde{\mathbf{A}} \alpha_{k}.
    \end{equation}
    After some manipulation, we find that a least-squares estimate for $ \tilde{\mathbf{A}} $ is given by
    \begin{equation} \label{eq:dmd-svd}
      \tilde{\mathbf{A}} = \mathbf{\Psi}_r^* \mathbf{X}' \mathbf{V}_r \boldsymbol{\Sigma}_r^{-1} \mathbf{\Psi}_r.
    \end{equation}

    For this anomaly detection application, the system dimensionality $ n $ will typically be much lower than the number of available time steps $ m,$ so the computation of $\mathbf{A}$ by equation \eqref{eq:dmd-pinv} is tractable.  However, we still use a dimensionality reduction approach, partly to take advantage of correlations in the time series and partly to make the method scalable to larger problems.  To estimate the full system matrix $ \mathbf{A} $ would require only a slight modification to equation \eqref{eq:dmd-svd}:
    \begin{equation}
      \mathbf{A} = \mathbf{X}' \mathbf{V}_r \boldsymbol{\Sigma}_r^{-1} \mathbf{\Psi}_r.
    \end{equation}

    Thus DMD provides an automated means of constructing an LTI model evolving the measurements of interest $\mathbf{x}$ in time, but we would also like this model to take into account readings from other sensors. Put another way, DMD finds the matrix $\mathbf{A}$ of equation \eqref{eq:kalman-filter} and must be extended to produce $\mathbf{B}$.

  \subsection{Dynamic Mode Decomposition with Control (DMDc)}
  \label{sec:dmdc}

    The full LTI system $ (\mathbf{A}, \mathbf{B}) $ can be estimated via a slight modification to the DMD procedure developed by Proctor, \textit{et al.} called {\em DMD with control} (DMDc)~\cite{Proctor2016siam}.  We split the full state vector into measurements of interest, $ \mathbf{x}, $ and exogenous ``inputs'' $ \mathbf{y} $. These inputs are treated as {\em actuation} in the Kalman filter model, but are more accurately taken to be simply exogenous predictors of the sensor measurements.  As with the state vectors, the input vectors can be compiled into a single matrix $ \mathbf{\Upsilon}. $  The estimation then proceeds similarly (as explained in detail in~\cite{Proctor2016siam}).  First data matrices are constructed
    \begin{align*}  
      \mathbf{X} = \begin{bmatrix}
        \vertbar & \vertbar & & \vertbar \\
        \mathbf{x}_1 & \mathbf{x}_2 & \cdots & \mathbf{x}_m\\
        \vertbar & \vertbar & & \vertbar 
      \end{bmatrix}, \hspace{0.5cm}
      \mathbf{X}' = \begin{bmatrix}
        \vertbar & \vertbar & & \vertbar \\
        \mathbf{x}_2 & \mathbf{x}_3 & \cdots & \mathbf{x}_{m+1}\\
        \vertbar & \vertbar & & \vertbar 
      \end{bmatrix},
      \hspace{0.5cm}
      \mathbf{\Upsilon} = \begin{bmatrix}
        \vertbar & \vertbar & & \vertbar \\
        \mathbf{y}_1 & \mathbf{y}_2 & \cdots & \mathbf{y}_{m}\\
        \vertbar & \vertbar & & \vertbar 
      \end{bmatrix}.
      \end{align*}
      The dynamics with control may be written in terms of these data matrices as
      \begin{align*}
      \mathbf{X}' \approx \mathbf{A} \mathbf{X} + \mathbf{B} \mathbf{\Upsilon}.
      \end{align*}
     Finally, the system matrices may be obtained by regression
     \begin{align*}
      \begin{bmatrix}
        \mathbf{A} & \mathbf{B}
      \end{bmatrix} = \mathbf{X}' \begin{bmatrix}
        \mathbf{X} \\ \mathbf{\Upsilon}
      \end{bmatrix}^\dagger.
    \end{align*}
    With the singular value decomposition
    \begin{equation}
      \begin{bmatrix}
        \mathbf{X} \\ \mathbf{\Upsilon}
      \end{bmatrix} = \begin{bmatrix}
        \mathbf{\Psi}_1 \\ \mathbf{\Psi}_2
      \end{bmatrix} \boldsymbol{\Sigma}  \boldsymbol{V}^*,
    \end{equation} the matrices $ \mathbf{\Psi}_1 $ and $ \mathbf{\Psi}_2 $ now give the principal components of the state and input subspaces, respectively.  Dimensionality reduction based on the singular values $ \boldsymbol{\Sigma}$ is also possible at this stage.

    The system $ (\mathbf{A}, \mathbf{B}) $ is finally estimated by
    \begin{align}
      \label{eq:A}
      \mathbf{A} &= \mathbf{X}' \boldsymbol{V} \boldsymbol{\Sigma}^{-1} \mathbf{\Psi}_1^*\\
      \mathbf{B} &= \mathbf{X}'  \boldsymbol{V}\boldsymbol{\Sigma}^{-1} \mathbf{\Psi}_2^*.
      \label{eq:B}
    \end{align}

  \subsection{Time delays: HAVOK and Delay-DMD}
  \label{sec:dmd-hankel}

    For complex dynamics, a standard linear system may not have enough descriptive ability to serve as a model for the Kalman filter fault detection method.
    One approach is to enrich the library with nonlinear functions of the state, leading to {\em Extended DMD} (EDMD) or {\em Koopman Mode Decomposition} (KMD)~\cite{williams_datadriven_2015}.
    EDMD/KMD has been demonstrated to yield models that are predictive enough for accurate full-state estimation with Kalman filters~\cite{Surana2016}, and for sensor fault detection in a power grid model~\cite{Surana2018jns}.
    In the latter work, anomalies were detected using hypothesis testing for the distribution of normalized innovation squared in the Kalman filter under the assumption of Gaussian white noise for process disturbances.
    However, accurate EDMD/KMD models rely on a judicious choice of observable functions~\cite{Brunton2016plosone,nathan2018applied}, which can be challenging in practice; the data matrices quickly become ill-conditioned as the number of observables is increased.

    For systems that cannot be accurately represented with standard DMD, we instead augment the library with time-delayed measurements.
    The use of time-delays has a long history in system identification, including the widely used Eigensystem Realization Algorithm and Observer Kalman Identification~\cite{ERA:1985, Juang1991} methods, and deep connections to dynamical systems theory~\cite{takens_detecting_1981, brunton_chaos_2017}.
    Augmenting the state vector with time-delays allows the model to capture some of the effects of latent variables.
    For example, consider a simple harmonic oscillator in periodic sinusoidal motion.
    A first-order one-dimensional linear model is only capable of expressing exponential growth and decay, not oscillatory dynamics.
    However, if the state is augmented by a time delay of 1/4 period, a linear model can effectively capture the second-order dynamics (or the latent, imaginary component of motion).
    Applying DMD to a time-delay-augmented vector can therefore give highly accurate representations of quasiperiodic dynamics~\cite{Champion2018}.

    The modeling and estimation procedure is effectively the same, except that a scalar measurement $x_k$  at time $t_k$ is replaced by a vector $\mathbf{x}_k = \begin{bmatrix}
    x_k & x_{k-d} & x_{k-2d} & \cdots & x_{k-n_d d}
    \end{bmatrix}^\top $, where $d$ is the length of each delay and $n_d$ is the number of delays.
    For anomaly detection applications, only the innovation corresponding to the current time step is tracked.

  \subsection{Decision trees}
  \label{sec:decision-tree}

    Decision trees are a popular machine learning method for both classification and regression problems~\cite{safavian1991survey,rokach2008data}.
    In this work we are interested in classification.
    Given a vector of real-valued features (e.g. sensor measurements or the innovation covariance), a decision tree uses a set of threshold-based rules to determine whether a failure has occurred. In this work the only features passed to the decision tree are raw sensor values and the average innovation covariance. However, it is common practice to engineer other features tailored to the specific problem domain.
    The rules are learned during a training phase in which labeled examples (the labels tell the tree which examples correspond to sensor failures) are shown to the tree.
    Once trained the rules are, in essence, a series of if-else statements with conditional expressions checking whether features are above or below different threshold values.
    We use the Scikit-learn decision tree implementation~\cite{scikit-learn}, trained with the CART algorithm~\cite{breiman1984classification}.

    Decision trees are a natural choice in this context because they automate and generalize the threshold-selection process.
    Rather than choosing, through trial and error, a cutoff in the innovation covariance above which a sensor failure is deemed to have occurred, a decision tree automates the choice.
    This approach also allows the model to take into account other sensor-to-sensor interactions not captured by the Kalman filter.
    In particular, the decision tree is able to set different innovation covariance thresholds for different regimes in the dynamics (e.g. perhaps one threshold is appropriate for low angle-of-attack maneuvers and another is better suited to high angle-of-attack flight patterns).
    This is especially important for the dataset of Section \ref{sec:real-world-dataset}.

    Decision trees have additional benefits relevant to our use-case: their decision mechanism is interpretable, they scale extremely well to large datasets and are fast to evaluate in an online setting, and they are able to identify which features are most useful.
    Without regularization they tend to overfit training data, so we limit the depth of the trees used in our experiments.

\section{The proposed method}
\label{sec:proposed-method}

  The following gives a high-level overview of the algorithmic method used for sensor failure detection.  We break the process into an offline phase where the model is calibrated using training data, 
  and an online phase where the model is deployed to detect sensor fault events in real time.

  \begin{enumerate}
    \item \textbf{Offline phase:} Given a training set $\mathcal{D}_{train}$ consisting of sensor measurements at various time points and corresponding labels,
    \begin{enumerate}
      \item Compute the DMDc system $(\mathbf{A},\mathbf{B})$ as in Section \ref{sec:dmdc} using time-delayed measurements described in Section \ref{sec:dmd-hankel};
      \item Derive any desired features from $\mathcal{D}_{train}$ to be used with the decision tree. This includes the average innovation covariance \eqref{eq:avg-innovation-cov}, which can be computed using $\mathbf{A}$ and $\mathbf{B}$;
      \item Train the decision tree using the derived and raw features.
    \end{enumerate}
    \item \textbf{Online phase:} Given a series of measurements $\x_1, \x_2,\dots$
    \begin{enumerate}
      \item Compute any features expected by the decision tree, using the previously constructed $\mathbf{A}$ and $\mathbf{B}$ to compute the average innovation covariance;
      \item Pass the features into the decision tree to obtain a class prediction (either that the sensor has failed or continues to function properly).
    \end{enumerate}
  \end{enumerate}

  The Kalman filter-based anomaly detection method described in this section can be applied to all available sensors simultaneously.
  The innovation covariance then becomes a matrix whose diagonal entries can be monitored to identify faults in the corresponding sensors.
  Note that in this work we restrict our attention to models for a single scalar state $x$.
  As such the decision tree is trained to output a binary result for a single sensor, although extension to parallel detection for multiple sensors is straightforward.

\section{Three example applications}
\label{sec:examples}

  \subsection{Real-world dataset}
  \label{sec:real-world-dataset}

\begin{figure}
      \centering
      \includegraphics[width=\textwidth]{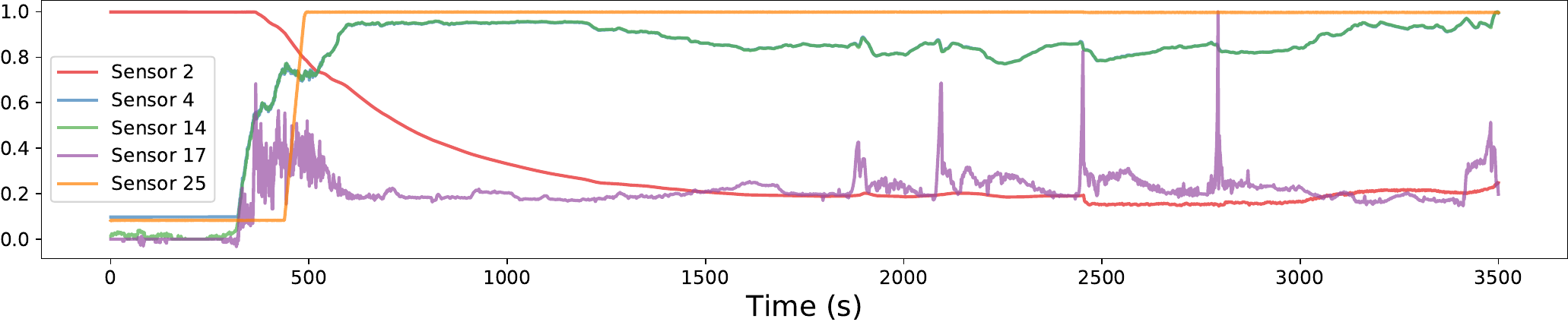}
      \caption{A plot of data from a subset of the sensors for test flight 1. Measurements from the faulty sensor (Sensor 2) are shown in red. The sensor fails just before second 2500, where there is a short drop followed by erratic, noisy measurements. We plotted maximally important features as returned by a decision tree trained on raw sensor data to predict failure events.}
      \label{fig:example-flight-data}
    \end{figure}

    We first consider the problem of detecting sensor failure using anonymized (scaled to lie in $[-1,1]$) measurements from aircraft sensors collected during flight tests. We focus on detecting faults in a given sensor that has a high failure rate (sensor 2) using the data from 25 other on-board sensors. Several sensors capture similar or redundant information which the DMD model can exploit. In total there are 21 flights, each roughly seven hours in length, with measurements recorded at a frequency of 20 Hz. Sensor 2 fails in 14 of the test flights. There is one time series in which the sensor failure was detected, sensor 2 was fixed, and then broke again. This case was split into two separate time series.
    When sensor 2 fails there is typically a small constant shift in the data followed by increased noise for the duration of the flight. Such a failure is shown along with anonymized data from some of the more relevant sensors in Fig.~\ref{fig:example-flight-data}.

    In this real data setting (data are from actual flights), the time of sensor failure time must be inferred from looking at the data.  
    We have hand-labeled estimated time of failure in each case; reported time to detection is based on these estimates.
    Systematic error in the absolute detection time is therefore possible, but comparisons between flight tests should be reliable.

    In order to train the decision tree the flights are divided into a training and testing set (no separate validation step is necessary as all parameters are determined with cross-validation using the training data). The training set consists of data from six test flights, flights 0 through 5, four of which contain a sensor failure.

    The DMD model is calibrated using flight 3, which contains no failure events. Notably, the model learns to predict future values of Sensor 2 by averaging together the current measurement from Sensor 2 with those of three other sensors with redundant signals (Sensors 0, 1, and 10). Each is given roughly equal weight. This is reminiscent of a weighted-mean method for consolidating redundant measurements into one fault tolerant estimate~\cite{allerton2005review}. From Fig. \ref{fig:example-flight-covariance} it is easy to see that something like the difference between sensors 2 and 10 would be a good proxy for when sensor 2 is behaving properly, but we stress that the model figures this out \textit{automatically}.

    A decision tree is then trained to predict sensor faults with its hyperparameters being selected using five-fold cross validation. Examples are reweighted before being fed to the tree in order to mitigate the effects of class imbalance. The top ten signals are given in Table \ref{tab:feat_weight_final} along with their Gini (feature) importances with respect to the decision tree.
    
    \begin{table}[t]
      \begin{center}
        \begin{tabular}{c c}
          \textbf{Feature} & \textbf{Importance}\\
          \hline
          $V_k$ & 0.7514\\
          Sensor 25 & 0.1985\\
          Sensor 4 & 0.0261  \\
          Sensor 5 & 0.0095  \\
          Sensor 9 & 0.0057  \\
          Sensor 14 & 0.0056 \\
          Sensor 17 & 0.0012 \\
          Sensor 22 & 0.0008 \\
          Sensor 13 & 0.0005 \\
          Sensor 7 & 0.0002  \\
        \end{tabular}
        \caption{Feature importance for the decision tree trained using the flight test dataset.}
        \label{tab:feat_weight_final}
        \end{center}
    \end{table}

    As seen in Table \ref{tab:feat_weight_final} most of the importance is concentrated in just two of the features, with the majority of the weight going to the moving average of the innovation covariance. The second-ranked feature, Sensor 25, turns out to be an almost binary signal indicating when measurements should be collected from the sensor of interest.

    The results of the detection of sensor failures on the flights in the test data set is summarized in Table~\ref{tab:results_fin}. We show performance metrics for each individual flight so that poor results can be more easily investigated. As there are so many data points for each flight, standard metrics such as precision (the proportion of flagged examples that were actually true positives) and recall (the proportion of true positives we were able to detect) are not particularly helpful here\footnote{True positives and true negatives are positive (failed sensor) and negative (working sensor) examples correctly classified by the model. False positives are negative examples the model classified as belonging to the positive class, i.e. instances where the sensors are working properly, but the model erroneously predicts a sensor has failed. Similarly, false negatives are positive examples the model thought were negative examples.}. Instead we focus on the number of false positives and false negatives along with lag time (time from actual sensor failure to detection). Overall accuracy is also included to give more context to the number of false positives and negatives reported.

    \begin{table}[t]
      \begin{center}
        \begin{tabular}{cccccc}
          \textbf{Flight} & \textbf{Total examples} & \textbf{False positives} & \textbf{False negatives} & \textbf{Accuracy} & \textbf{Lag Time (s)}\\
          \hline
          6   & 175,162 &    708 &      0 &  0.995958 &      N/A \\
          7   & 213,964 &      0 &     22 &  0.999897 &      1.1 \\
          8   & 143,140 &      0 &  4,663 &  0.967424 &     10.1 \\
          9   & 121,413 &      0 &    412 &  0.996607 &     1.35 \\
          10  & 368,140 &      0 &      0 &  1.000000 &      N/A \\
          11  & 152,146 &      0 &    130 &  0.999146 &      6.5 \\
          12  & 278,465 &     50 &      0 &  0.999820 &      N/A \\
          13  & 372,950 &      3 &      0 &  0.999992 &      N/A \\
          14  & 124,570 &  2,617 &      0 &  0.978992 &      N/A \\
          15  & 166,880 &      0 &      0 &  1.000000 &        0 \\
          16  & 302,550 &      0 & 11,690 &  0.961362 & $\infty$ \\
          17  & 295,680 &      0 &     63 &  0.999787 &      2.9 \\
          18  &  64,700 &      0 &      9 &  0.999861 &     0.45 \\
          19  & 472,090 & 59,809 &    157 &  0.872978 &     7.85 \\
          20  & 117,650 &      0 &    229 &  0.998054 &    11.45 \\
        \end{tabular}
        \caption{Prediction results for commercial test flights.}
        \label{tab:results_fin}
      \end{center}
    \end{table} 

    Assuming an end goal of using our model to alert a human of potential sensor failure during a flight test within a few seconds of occurence, the model performance is promising, with the exception of two flights: 16 and 19.
    There is some fluctuation in lag time, most likely due to circumstances that affect flight dynamics.
    The false negatives are almost entirely due to lag time.
    The false positives tend to persist briefly (about two seconds) before automatically correcting. 
    For example for flight 8 we observe four different short periods when the model thinks a sensor failure has occurred. These correspond to abrupt changes in flight conditions as the pilot(s) carry out different test maneuvers.

    Now we turn our attention to the two flights with poor model performance.
    In flight 16 there are a large number of false negatives (and a lag time of $\infty$) as the model completely misses an actual sensor failure.
    This is because sensor 25, which is normally active during the tests, was completely inactive throughout the entirety of this flight.
    If one wished to learn to predict regardless of whether or not this sensor was in use, one may need to resort to training a separate model using only data where the sensor was inactive, because it is used in all the other flights.

    In flight 19 we see a large number of false positives.
    This is caused by persistent discrepancies between sensor 2 and its redundant sibling, sensor 10. We suspect that sensor 10 temporarily malfunctioned.
    Whether or not it is desirable for the algorithm to flag these anomalous measurements depends on the application.
    After the abnormal behavior of sensor 10 the model detects the failure of sensor 2 almost perfectly.

    The final model is able to reliably detect true sensor failure events within 12 seconds (with most occurring in under five). 
    There are infrequent false positives detected with a brief persistence of about two seconds. In some cases the measurements taken by the faulty sensor drift close to the values of the reliable redundant sensors, leading to false negatives.

    \begin{figure}
      \centering
      \includegraphics[width=.75\textwidth]{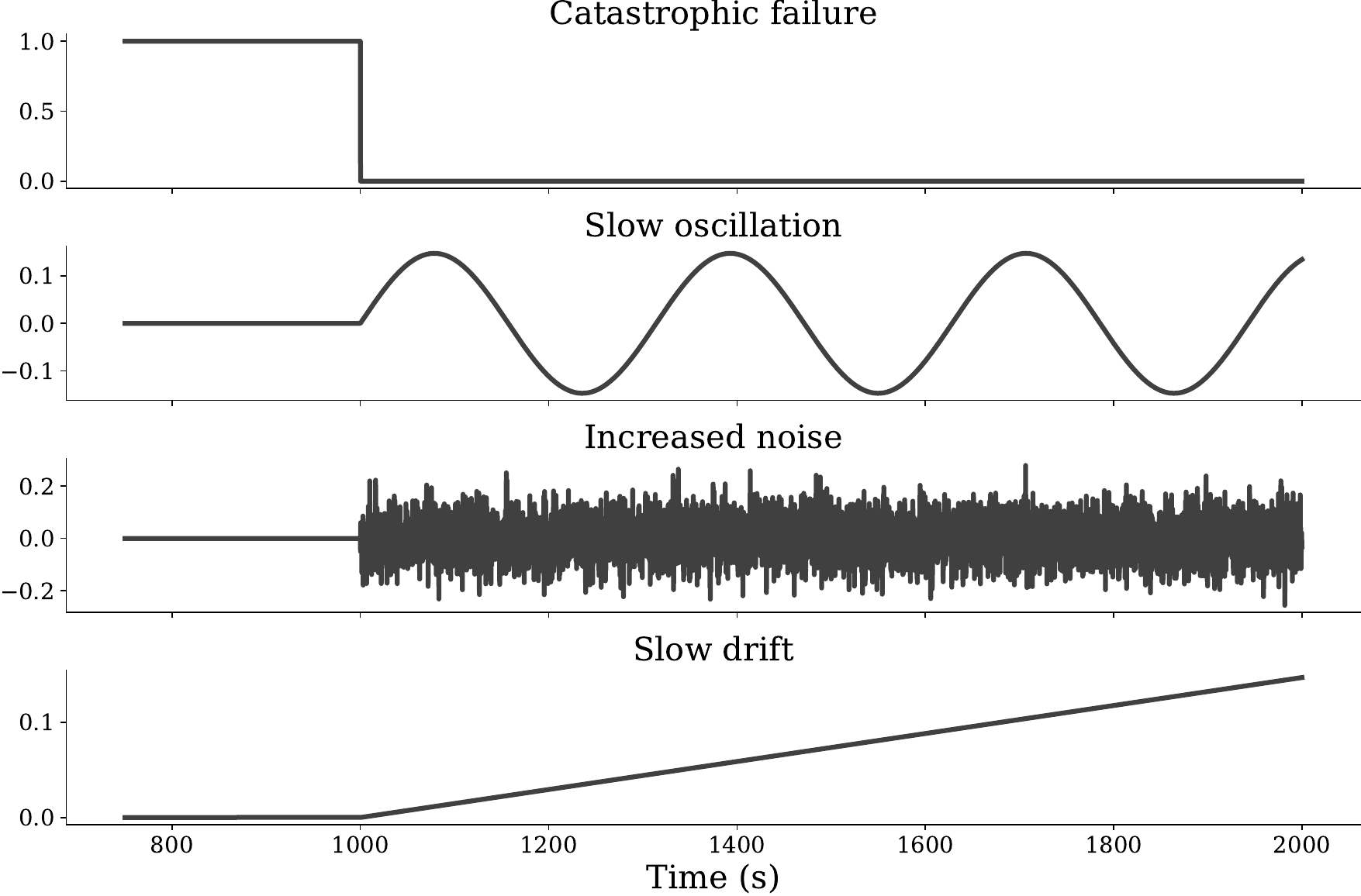}
      \caption{The sensor faults applied to the simulated datasets. In particular, the faults used for the dataset of Section \ref{sec:goman-khrabrov} are shown here. Those used for the dataset of Section \ref{sec:flight-dynamics-model} are the same up to rescaling.}
      \label{fig:sensor-failure-modes}
    \end{figure}

  \subsection{Synthetic datasets}
  \label{sec:synthetic-datasets}

    In order to better understand the capabilities of our proposed approach we apply it to two sensor failure tasks derived from simulated datasets. In both instances we use a dynamical model to generate realistic sensor data, which we then augment in various ways to mimic common sensor failure modes. Specifically, we simulate the faults shown in Fig. \ref{fig:sensor-failure-modes}, similar to those studied by Eykeren and Chu~\cite{vanEykeren2014}: catastrophic failure (multiplicative), slow oscillation (additive), increased noise (additive), and slow drift (additive). In every case we add a small amount of white noise (mean 0, standard deviation $5\times 10^{-3}$) to the underlying sensor measurement of interest \textit{after} incorporating the fault.

    \subsubsection{Goman-Khrabrov model}
    \label{sec:goman-khrabrov}

      A long-standing challenge in aerodynamic modeling was capturing the effect of a separated flow on aerodynamic moments.
      This arises for example in high angle-of-attack maneuvers, when an airfoil wake can detach, leading to much more complex physical behavior such as dynamic stall \cite{Leishman2002book}.  
      The aerodynamic moments in this case depend not only on the airfoil configuration, as in standard linearized approaches based on stability and control derivatives, but also on the state of the separated flow. 
      Because of their importance in a range of unsteady fluid dynamic contexts, dynamic stall, and separated flows more generally, have been widely studied~\cite{Seifert:JA96,amitay2002controlled,magill:03dsv,taira:08tip,canonical:2010,colonius2011control,Dunne2015ef,an2016modeling,Taira2017aiaa,Eldredge2019arfm}.

      Goman and Khrabrov proposed a mathematical model of dynamic stall that treats the flow state as a dynamic internal system variable \cite{Goman1994ja}.
      For the case of a high angle-of-attack airfoil this is a scalar variable $x \in (0, 1)$ representing the separation point normalized by the chord length, so that fully attached flow corresponds to $x=1$.
      The internal flow field dynamics are modeled with a simple time-delay model:
      \begin{equation}
        \label{eq:gk-internal}
        \tau_2 \dot{x} + x = x_0\left(\alpha - \tau_1 \dot{\alpha}\right).
      \end{equation}
      The function $x_0(\alpha)$ defines the empirical steady separation point as a function of angle of attack $\alpha$.
      Quasisteady effects are expressed through the time-delay shift $\tau_1 \dot{\alpha}$.
      The overall model dynamics are a relaxation towards the quasisteady separation point on a timescale $\tau_2$.
      The moments are then algebraic functions of the aerodynamic state, for example $C_L = C_L(\alpha, x) $.
      For a high angle-of-attack airfoil, the model
      \begin{equation}
        C_L(\alpha, x) = \frac{\pi}{2} \sin \left[ \alpha (1 + \sqrt{x})^2 \right]
      \end{equation}
      was shown to accurately describe experimental data for a NACA 0015 airfoil \cite{Goman1994ja}.

      For our synthetic data, we use a simple model for the steady separation point:
      \begin{equation}
        x_0(\alpha) = \frac{1 - \tanh\left[20(\alpha - 0.25)\right]}{2}.
      \end{equation}
      This produces the expected hysteretic behavior, although it is not expected to accurately represent the steady separation point of any particular airfoil.
      The time constants $\tau_1$ and $\tau_2$ in the model can be obtained in general by fitting to experimental data. 
      We use the reported values $\tau_1 = 0.5$ and $\tau_2 = 4.5$, nondimensionalized by chord length and free stream velocity \cite{Goman1994ja}.

\begin{figure}
      \vspace{-.1in}
        \centering
        \includegraphics[width=0.6\textwidth]{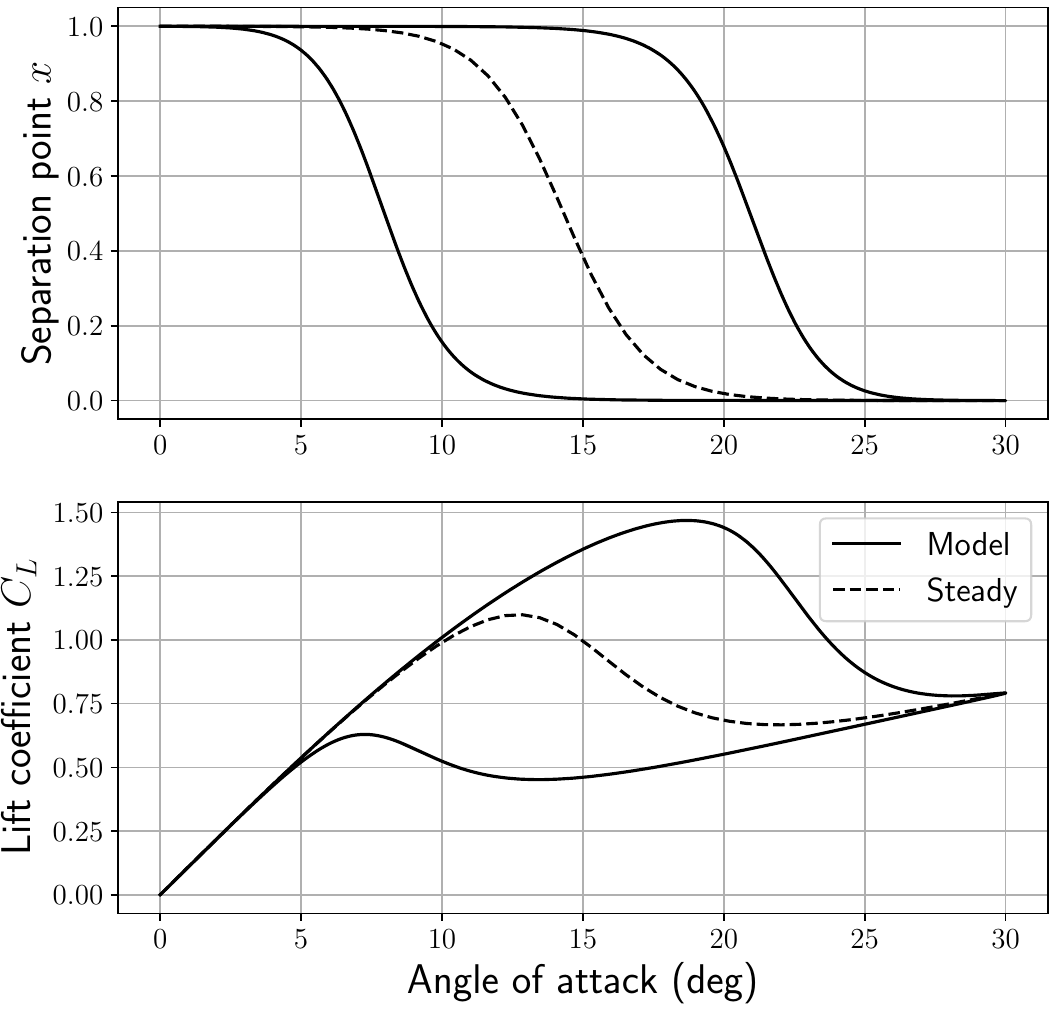}
        \caption{Hysteresis in flow separation (top) and lift coefficient (bottom) in the Goman-Khrabrov model for an airfoil undergoing sinusoidal pitching motion at nondimensional frequency $\omega = 0.05$. Stall is delayed relative to the steady value for pitch-up motions with attached flow (upper curves on both plots).}
        \label{fig:gk-AoA}
      \end{figure}

      The separation point and lift coefficient for a sinusoidal pitching motion at nondimensional frequency $\omega = 0.05$ is shown in figure \ref{fig:gk-AoA}, along with the steady values as a function of angle of attack.
      The effect of the model is to capture observed hysteresis in the curves; the flow remains attached to higher angle-of-attack on pitch-up motions and stall is delayed.
      Conversely, when the flow is separated during a pitch-down maneuver it remains so for longer, resulting in reduced lift relative to the steady value.

      \begin{figure}
      \vspace{-.1in}
        \centering
        \begin{subfigure}{\textwidth}
          \centering
          \includegraphics[width=.8\textwidth]{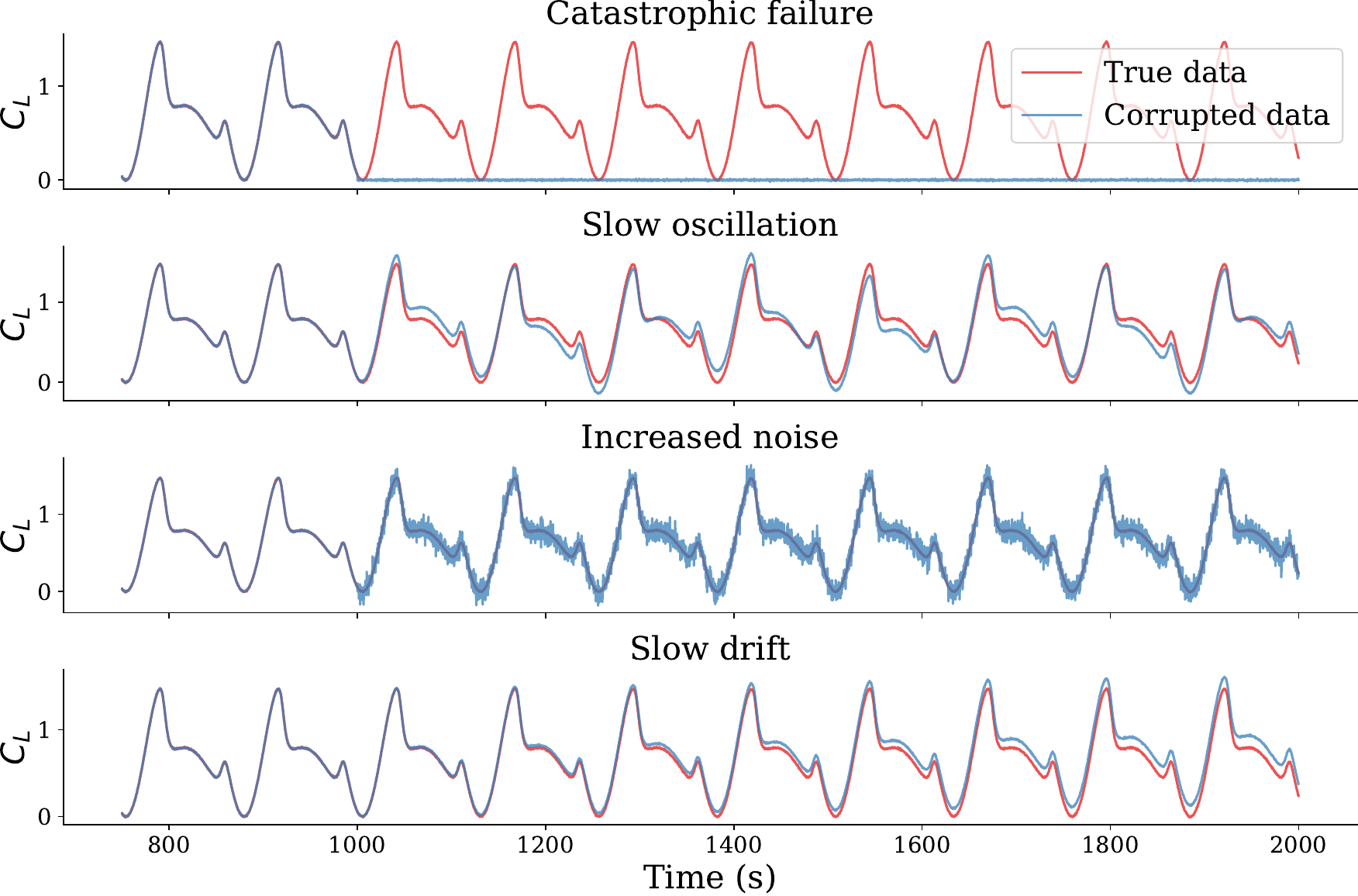}
          \caption{Simulated data}
          \label{fig:goman-khrabrov-data}
        \end{subfigure}%
        \\
        \vspace{.4in}
        \begin{subfigure}{\textwidth}
          \centering
          \includegraphics[width=.8\textwidth]{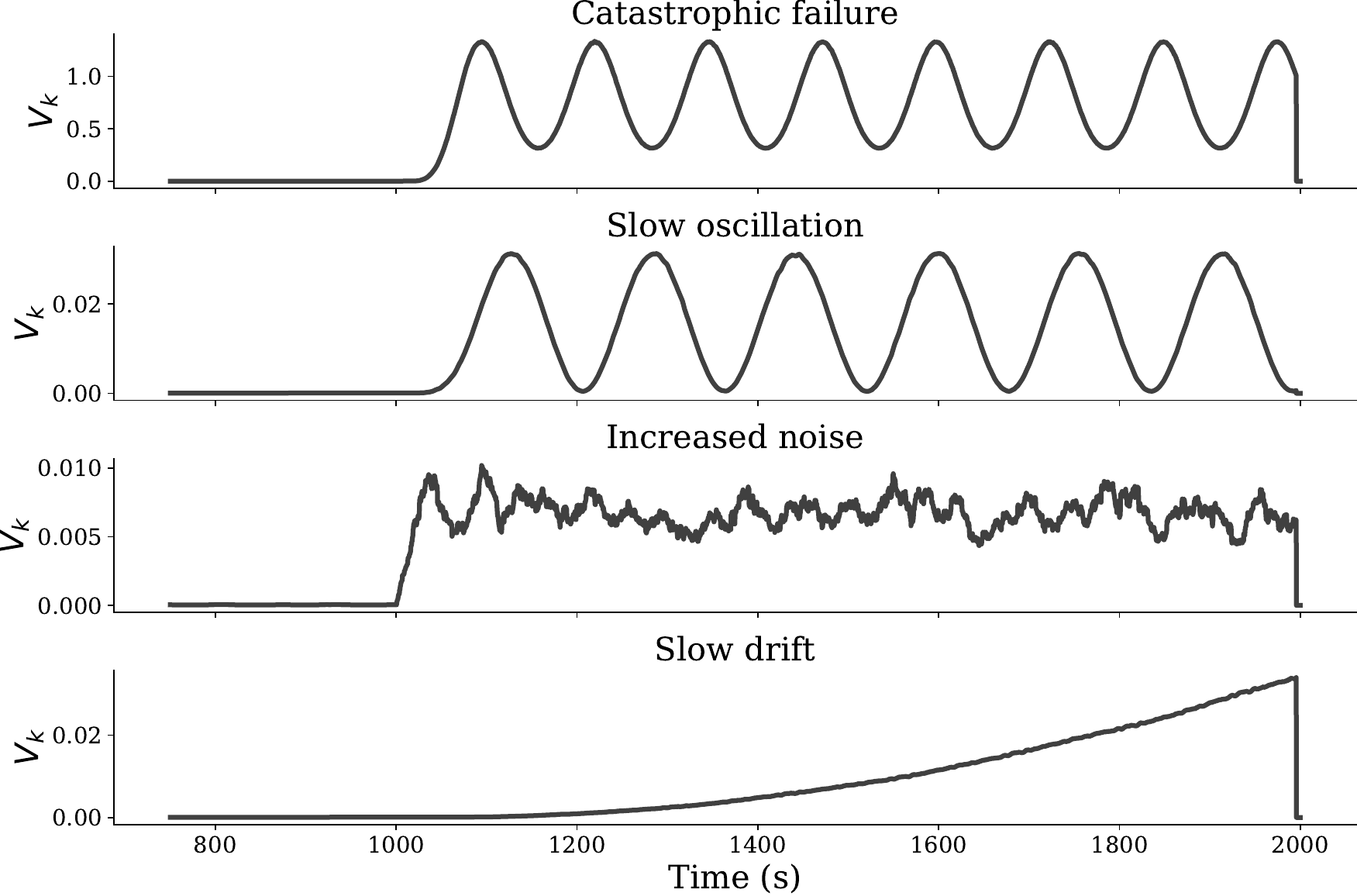}
          \caption{Innovation covariance}
          \label{fig:goman-khrabrov-cov}
        \end{subfigure}
        \caption{(a) Visualizations of the different lift coefficient ($C_L$) sensor failure modes for the Goman-Khrabrov model. (b) Moving average of the innovation covariance for each sensor fault type using data generated with the Goman-Khrabrov model. Sensor failure occurs at $t=1000$. Note that we omit from this plot measurements taken with $t<750$.}
      \end{figure}

      Our experimental dataset consists of measurements of $C_L$, $\alpha$, and $\dot \alpha$ taken five times per second for 2000 seconds. The DMDc model is trained using the full time-series. We then simulate failure of a hypothetical $C_L$ sensor at $t=1000$ using each of the fault modes shown in Fig. \ref{fig:sensor-failure-modes}, which results in the time series given in Fig. \ref{fig:goman-khrabrov-data}. We then precompute the innovation covariance, $V_k$, for each sensor failure type, with the Kalman filter \eqref{eq:kalman-filter} constructed using DMDc. The scale of $V_k$ differs between the modes so only training on one type of sensor failure may lead to poor results when trying to detect different failure types.
      We plot the moving average of the covariance for the different failure modes in Fig. \ref{fig:goman-khrabrov-cov}. 
      Note the differences in scale between the modes. A threshold for $V_k$, above which a sensor failure is deemed to have occurred, which is chosen based on only one type of fault, may be inappropriate for the others.

      Finally, we train a decision tree to predict when a sensor failure has occurred.
      The tree is given access to $\alpha$, $\dot\alpha$, the corrupted $C_L$ measurements, and $V_k$ as features.
      At each time point the model must attempt to predict whether the given $C_L$ measurement has been corrupted or not (whether the sensor has failed).
      We use five-fold cross-validation to select model parameters.
      We construct two types of training and testing data;
      the first involves incorporating random examples from \textit{all four} failure types into the training set and the second sources its training data from just one sensor fault and attempts to predict when the others have occurred. In each case the training set contains 30,000 examples and the testing set contains 10,000 examples.

      Model performance is summarized in Table \ref{tab:gk-metrics}. ``Train accuracy'' refers to the accuracy of the model on a holdout set during cross-validation. For models 1-4 this number gives an estimate of the models' accuracies on the fault type on which they were trained. The best accuracy and precision scores are both achieved by the model with training data from all four fault types. However, model 4 has the best recall, meaning that it misses the fewest sensor failure events. This is likely due to the fact that model 4 must choose a very small threshold for $V_k$ at which to separate negative and positive class instances.
      All five models list the average innovation covariance as their top feature.

      \begin{table}[t]
        \centering
        \begin{tabular}{ccccccc}
          \textbf{Model} & \textbf{Fault types seen} & \textbf{Accuracy} & \textbf{Precision} & \textbf{Recall} & \textbf{Train accuracy} & \textbf{Tree depth} \\ \hline
          1 & Catastrophic failure & 0.9732 & 0.9980 & 0.9483 & 0.9343 & 5\\
          2 & Slow oscillation & 0.9672 & 0.9629 & 0.9718 & 0.9351 & \textbf{2}\\
          3 & Increased noise & 0.9730 & 0.9993 & 0.9467 & 0.9482 & \textbf{2}\\
          4 & Slow drift & 0.9784 & 0.9707 & \textbf{0.9867} & 0.8678 & 3\\
          5 & All & \textbf{0.9818} & \textbf{0.9996} & 0.9642 & \textbf{0.9823} & 5\\
        \end{tabular}
        \caption{Performance metrics for models trained on different subsets of sensor fault types with data generated with the Golman-Khrabrov model. The best values for each column are bolded.}
        \label{tab:gk-metrics}
      \end{table}

      The proposed method reliably detects sensor faults for data generated from the Goman-Khrabrov model, even on unseen fault types. 
      However, if either the amplitude or frequency of the forcing term changes after the DMD model has already been trained, the DMD model becomes too inaccurate to be useful and overall predictive performance suffers considerably. This is a general drawback of data-driven models: when training and testing sets are different enough in distribution, models learned on one set have a hard time generalizing to the other.

    \subsubsection{Flight dynamics model}
    \label{sec:flight-dynamics-model}

      Motivated by anomaly detection in a flight test setting, we also consider a longitudinal flight dynamics model for a business jet in atmospheric turbulence \cite{Stengel2004book}.
      The equations of motion for the longitudinal model capture motions in the forward and vertical directions, including pitch for a total of three degrees of freedom.
      Flight controls are included for elevator, thrust, flaps, and stabilator; these are trimmed for steady, level flight.
      The aerodynamic model includes a realistic geometric configuration, stability and control derivatives, US standard atmospheric conditions interpolation, and a Mach number correction.
      The dynamics are forced by atmospheric turbulence generated to approximate the von K\`{a}rm\`{a}n spectrum by filtering band-limited white noise \cite{MIL-HDBK-1797, Wang1980nasa}.

\begin{figure}
      \vspace{-.1in}
        \centering
        \begin{subfigure}{\textwidth}
        \centering
          \includegraphics[width=.8\textwidth]{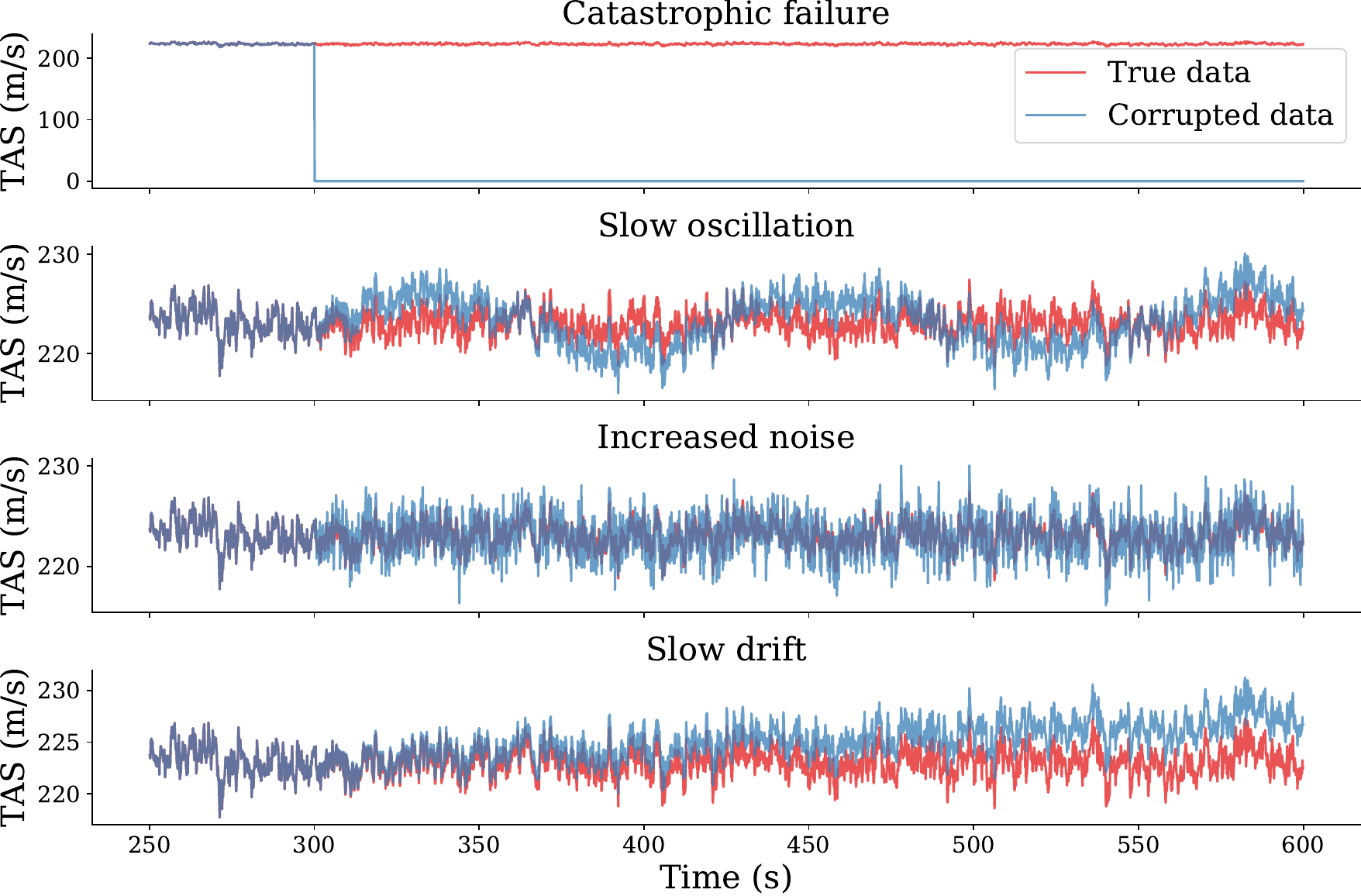}
          \caption{Simulated data}
          \label{fig:longitudinal-model-data}
        \end{subfigure}
        \\  
        \vspace{.5in}
        \begin{subfigure}{\textwidth}
          \centering
          \includegraphics[width=.8\textwidth]{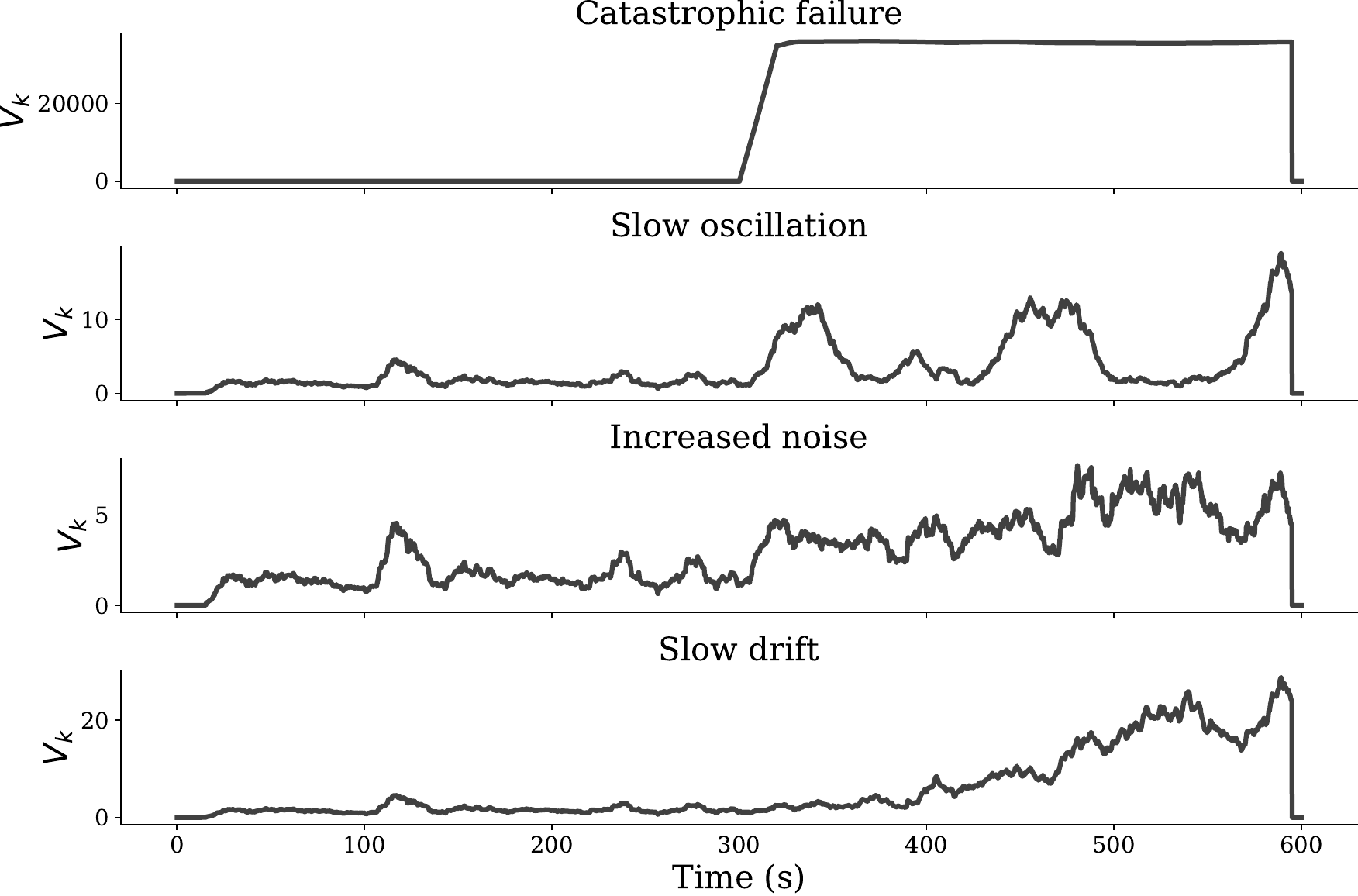}
          \caption{Average innovation covariance} 
          \label{fig:longitudinal-model-cov}
        \end{subfigure}
        \caption{Flight dynamics model data. (a) Visualizations of the different true airspeed (TAS) sensor failure modes. (b) Moving average of the innovation covariance for each sensor fault type. Sensor failure occurs at $t=300$. Note that we omit from (a) measurements taken with $t<250$. We retain them in (b) to show the behavior of the innovation covariance for normal sensor readings.}
      \end{figure}

      From this system we can measure not only the dynamic variables for inertial velocity and pitch, but also lift, drag, pitching moment, true airspeed, angle of attack, and Mach number.
      We consider measurements of the true airspeed (TAS), informed also by angle of attack, inertial airspeed, pitch, lift, and thrust.
      Note that the flight controls are constant, but the model includes an altitude correction for effective thrust.
      We generate samples for each variable over $t\in[0, 600]$ at a rate of 10 samples per second. Next we build a Kalman filter from a DMDc model trained on the TAS data. Various faults are then introduced starting at $t=300$ in the TAS sensor to obtain the time series shown in Fig. \ref{fig:longitudinal-model-data}. These measurements are much noisier than those from the previous section due to the turbulence-based forcing.

        Again we precompute the innovation covariance, $V_k$, for each fault type. The results are shown in Fig. \ref{fig:longitudinal-model-cov}. It is evident from this plot that all faults, except the catastrophic failure, will be difficult to detect. None of the other covariance time series admit a single threshold that distinguishes bad readings from good ones.
      As before we study the behavior of decision trees trained on data from all four failure modes or from just one mode. The trees are given measurements of TAS, angle of attack, inertial air speed, pitch, lift, thrust, and $V_k$. Training and testing sets are of equal size, with each training set constituting 75\% of the data. Since the dataset is more complex we allow our cross-validation procedure to select trees of depth up to seven.

      Table \ref{tab:fd-metrics} details the results of these experiments.
      Model 5 outperforms all the others by a considerable margin, though it is also the most complicated model.
      Since the decision tree sees examples of each failure type, it is able to construct a more complex set of thresholding rules for different situations.
      In contrast, the other models consist of trees which appear to \textit{overfit} to the failure modes they are shown, as evidenced by their diminished accuracy scores relative to model 5.
      Notably, model 1 adopts a classification rule allowing it to predict catastrophic sensor failures with perfect accuracy (its in-group accuracy is 1), however for all other failure types it is no better than a coin toss.
      The threshold it selects for $V_k$ is much too large for the other fault types and so it almost always predicts that no sensor failure has occurred.
      In this case model 2 has the highest recall, but also a low precision, for the same reason as model 4 did previously: its threshold for $V_k$ is set lower than the others, reducing the number of false negatives at the cost of more false positives.
      The average innovation covariance has the highest feature importance for each of the five models.
      \begin{table}[t]
        \centering
        \begin{tabular}{ccccccc}
          \textbf{Model} & \textbf{Fault types seen} & \textbf{Accuracy} & \textbf{Precision} & \textbf{Recall} & \textbf{Train accuracy} & \textbf{Tree depth} \\  \hline
          1 & Catastrophic failure &  0.5002 & 0 & 0 & 1.0 & \textbf{2} \\
          2 & Slow oscillation     &  0.9549 & 0.9230 & \textbf{0.9926} & 0.8412 & 3 \\
          3 & Increased noise      &  0.8974 & 0.9915 & 0.8016 & 0.8878 & \textbf{2} \\
          4 & Slow drift           &  0.9572 & 0.9659 & 0.9478 & 0.8190 & 4 \\
          5 & All                  &  \textbf{0.9867} & \textbf{0.9961} & 0.9767 & \textbf{0.9859} & 7 \\
        \end{tabular}
        \caption{Performance metrics for models trained on different subsets of sensor fault types for the flight dynamics model. The best values for each column are shown in bold.}
        \label{tab:fd-metrics}
      \end{table}

      The proposed method is effective at identifying sensor faults in simulated flight test data, but there is a noticeable degradation in performance relative to data simulated with the Goman-Khrabrov model. This is due in part to the complexity of the simulated dynamics as well as the erratic atmospheric forcing.
      We observe that for this more challenging dataset, it is increasingly important that the model be trained using examples from multiple sensor failure modes, allowing it to establish different decision thresholds for different types of dynamics.

      If data for some failure types is unobtainable, then the model should be trained using fault types that are most difficult to detect.
        Failure modes that can be detected trivially, such as catastrophic sensor failure, may prove insufficient to train a robust detector.

\section{Conclusion}
\label{sec:conclusion}
  
  We have developed a fully automatic approach to detect sensor failures  in systems with multiple types of sensor failures.
  The method first uses the dynamic mode decomposition for control with time-delay measurements to learn a simple linear time-invariant model for the evolution of a sensor of interest in time.
  This model is embedded in a Kalman observer which is then used to predict future measurements.
  A potential sensor fault is detected when the predicted and measured sensor values disagree by a large margin, with the margin size selected using a decision tree. All components are trained automatically.
  The performance of the proposed method was demonstrated on three test datasets: real measurements from a series of flight tests and two simulated datasets from the Goman-Khrabrov and a realistic flight dynamics model.
  In each case the difference between the true and Kalman-observer-predicted values of the sensor of interest provided an accurate proxy for when sensor failure had occurred.

  There are numerous extensions that could be explored for improving upon the results obtained here. Any of the components of the algorithm could be replaced with more sophisticated variants. For example, advances in Koopman theory could be leveraged to enrich the linear time invariant physics model.
  A nonlinear model such as an extended Kalman filter or a more general estimator~\cite{aravkin2017generalized} or a model learned via some other model discovery framework \cite{Schmidt2009science,brunton_discovering_2016,Kaiser2018prsa} could be used in place of the Kalman filter.
  Such generalizations would allow for the application of the proposed method to systems exhibiting strongly nonlinear dynamics.
  It would also be interesting to incorporate this analysis within the context of robust statistics~\cite{Candes:2011}, which has recently been shown to improve flow measurements~\cite{Scherl2019arxiv}. 
  The performance of the decision tree could be enhanced by employing an ensemble \cite{liaw2002classification}, a cost-sensitive training algorithm \cite{lomax2013survey}, or by better utilizing class probabilities output by the tree.

\section*{Funding Sources}
  The authors acknowledge support from the Boeing Company (award 2018-ETT-PA-379). 
  JNK also acknowledges support from the Air Force Office of Scientific Research (FA9550-19-1-0011). 
  SLB also acknowledges support from the Air Force Office of Scientific Research (FA9550-18-1-0200). 

\bibliographystyle{plain}
\begin{spacing}{.9}
  \small{
    \setlength{\bibsep}{5.pt}
    \bibliographystyle{unsrt}
    \bibliography{anomaly_paper}
  }
\end{spacing}

\end{document}